\documentclass{aa}
\usepackage[english]{babel}
\usepackage{psfig,graphicx}
\usepackage{supertab}

\onecolumn
\begin{document}

\title{IR-source IRAS\,20508+2011: spectral variability of the optical
         component}

\author{V.G.~Klochkova$^1$, V.E.~Panchuk$^1$, N.S.~Tavolganskaya$^1$, and G.~Zhao$^2$}

\date{\today}	     

\institute{Special Astrophysical Observatory RAS, Nizhnij Arkhyz,  369167 Russia, 
\and     National Astronomical Observatories CAS,  Beijing, China}

\abstract{Based on high-resolution spectra we revealed variability of the optical
spectrum of the cool star identified with the IR source IRAS\,20508+2011.
Over the five years of our observations, the radial velocity derived from
photospheric absorption lines varied in the interval $V_r = 15-30$\,km/s.
In the same time, the H$\alpha$ profile varied from an intense bell-shaped
emission line with a small absorption to 2-peaked emission with a central
absorption feature below the continuum level. At al but one epoch, the
positions of the metallic photospheric lines were systematically shifted
relative to the H$\alpha$ emission: $\Delta V_r=Vr(met)- V_r(H,
emis)\approx -23$\,km/s. The Na\,D doublet lines shown a complex profile
with broad (half-width $\approx$\,120\,km/s) emission and photospheric
absorption, as well as an interstellar component. We used model
atmospheres to determine the physical parameters and chemical composition
of the star's atmosphere: $T_{eff}$=4800\,K, $\log g$=1.5,
$\xi_t$=4.0\,km/s and metallicity [Fe/H]=-0.36. We detected overabundances
of oxygen [O/Fe] =+1.79 (with the ratio [C/O]$\approx$-0.9), and
$\alpha$-process elements, as well as a deficit of heavy metals. The
totality of the parameters suggests that the optical component of
IRAS\,20508+2011 is an ``O-rich'' AGB star with luminosity
$M_v\approx-3^m$ that is close to its evolution transition to the post-AGB
stage.
}

\authorrunning{Klochkova et al.}
\titlerunning{Spectroscopy of IRAS20508+2011}

\maketitle

\section{Introduction}

Our program of spectroscopic observations of evolved stars identified with
IR sources is being carried out with the 6-m telescope of the Special
Astrophysical Observatory of the Russian Academy of Sciences. Its basic
goal is to determine the main parameters and evolutionary stages of the
stars, search for evolutionary variations of their chemical compositions,
study in detail the velocity fields in their atmospheres and envelopes,
and detect spectral variability and binarity. The objects under
consideration are basically intermediate-mass stars in the
asymptotic-giant-branch (AGB) and postAGB stages, which have evolved
through stages of hydrogen and helium burning both in the core and layer
sources, mixing and ejection of matter, and mass loss via the stellar wind
on the AGB. There are strong grounds to expect variable elemental
abundances in atmospheres of these evolved objects. Stars in the
transition from the AGB to planetary nebulae are conventionally called
protoplanetary nebulae (PPN). The main results of the program are briefly
summarized by Klochkova et al. [\cite{KPS}]. The papers [\cite{Egg}],
which is the first in our series of works on the spectroscopy of bipolar
nebulae, and [\cite{IRAS01005}], which presents the first detailed
spectroscopic results for the post-AGB object IRAS\,01005+7910 are also
important. The peculiar spectral variability of this object is apparently
due to inhomogeneity of its circumstellar gas-dust envelope. We present
here the results of our first spectral observations of the optical
component of the IR source IRAS\,20508+2011 (hereafter IRAS\,20508). The
source is beyond the Galactic plane ($b = -15\lefteqn{.}^o3$, $l=
66\lefteqn{.}^o0$), possibly indicating that it does not belong to the
young disk population. The IRAS catalog [\cite{IRAS}] presents the IR
fluxes for the source $f_{12}$=0.71, $f_{25}$=1.10, $f_{60}$=0.48, and
$f_{100}$=1.00.

Unfortunately, neither multicolor photometric nor spectral data at optical
wavelengths that could be used to derive the parameters of the object and
reveal its evolutionary status are available. Using the interactive ALADIN
atlas from the Strasburg CDS database, we identified the optical component
of IRAS\,20508 as the star No.\,815697 from the compiled catalog
[\cite{Khar}], with $B=12\lefteqn{.}^m868$ and $V=11\lefteqn{.}^m938$
(color index $B-V=0\lefteqn{.}^m93$). The star is also present in the
catalog GSC HST [\cite{HST}]: No.\,0165500558
($mph=11\lefteqn{.}^m6$). Detection of strong H$\alpha$ emission in the
first spectrum of the star obtained with the 6-m telescope in summer 1999
led us to continue the observations. In Sect.\.2, we give a brief description
our observations and the processing of the spectral data. Sect.\,3 describes
the features we found in the spectra of the optical component of IRAS\,20508,
as well as our determination of its chemical composition and the radial
velocities in the stellar atmosphere and the circumstellar material.

\section{Observations and data reduction}

We carried out spectroscopic observations of IRAS\,20508 with the 6\,m
telescope of the SAO. Three spectra were obtained at the prime focus with
the PFES echelle spectrograph [\cite{PFES}], which provides a spectral
resolution of R$\approx$17000 in combination with a 1160$\times$1040 CCD.
The two latest spectra were obtained with the NES echelle spectrograph
[\cite{nes}] at the Nasmyth focus, using a 2048$\times$2048 CCD. The NES
spectrograph equipped with an image slicer [\cite{slicer}] provides a
spectral resolution R=60000. Table~1 presents the moments of 
observations and the wavelength region of the spectra. The data were
extracted from the two-dimensional echelle spectra using the ECHELLE
procedure in the MIDAS package. Cosmic-ray traces were removed via median
averaging of two spectra obtained consecutively. The wavelengths were
calibrated using spectra from a Th-Ar hollow-cathode lamp. The further
processing, including photometric and position measurements, was carried
out with the DECH20 code [\cite{gala}], which enables the determination of
positions for individual spectral features by bringing into coincidence
the direct and mirror images of their profiles. To increase the accuracy
of our radial-velocity measurements, we selected unblended lines for
comparisons between observed and synthetic spectra. For each spectrogram,
the positional zero point was determined in the standard way, by
calibrating based on the positions of ionosphere night sky emission and
telluric absorption lines observed against the background of the object's
spectrum. The accuracy of the velocity measured from a single line in a
spectrum obtained with the PFES spectrograph is about 3.0\,km/s;
the NES spectrograph provifed to measure such single-line velocities with
an accuracy of about 1\,km/s.

\section{Discussion}

\subsection{Radial velocity pattern of IRAS\,20508}

In whole the optical spectrum of IRAS\,20508 is typical of a late-type
supergiant. Fig.\,1 compares a part of the observed spectrum with the
theoretical spectrum calculated in the LTE--approximation with the model
parameters $T_{eff}$=4800\,K, $\log g$=1.5, and $\xi_t$=4.0\,km/s (the
method and accuracy of the model parameters are discussed below, in
Subsect.\,3.2). No signs of the molecular bands that are often observed in
the optical spectra of PPN (see, for example, [\cite{KPS, Bakk97,
IRAS08005}]) are visible in the spectrum of\,IRAS 20508. The entire set of
radial velocities derived from individual lines did not reveal any
wavelength dependence of the velocity. Table\,2 presents the measurements
of the radial velocity $V_r$ for all dates. This table indicates essential
time variability of $V_r$ averaged over the lines of metals. At the epoch
of our observations, the amplitude of the $V_r$ variations measured from
metallic absorption lines is about 15\,km/s: $V_r$=15.5--29.9\,km/s.
Within the uncertainties, the positions of the absorption cores of the H$\beta$ 
and H$\gamma$ lines whose profiles are not distorted by emission are consistent
with the velocity derived from the metallic lines. It follows from Table\,2
that the velocity corresponding to the H$\alpha$ emission wings is variable, which
may indicate the presence of some disturbing component in the IRAS\,20508
system. Further observations are needed to confirm the binarity of the
star.

\subsubsection{The H$\alpha$ profile variability.}

The main feature in the optical spectrum of IRAS\,20508 is the intense
H$\alpha$ emission (Fig.\,2). Since the H profiles in both spectra
obtained in 2000 are very similar, only one profile is plotted in Fig.\,2
for this year. H$\alpha$ emission is common for stars evolving close to
the AGB stage and, along with an IR excess, provides the main criterion
for PPN candidates [13]. In the spectra of typical PPN, the H$\alpha$ line
displays complex (emission + absorption) variable profiles with core
asymmetry, P~Cygni or inverse P~Cygni profiles, or profiles with
asymmetrical emission wings (see, for example, [\cite{rev, Maas}]). Such features
are also frequently observed in combinations with each other. H$\alpha$ emission
is known as a sign of matter outflow and/or pulsations. In the case of
PPN, as a rule, the shift of the core is smaller than that corresponding
to the escape velocity; i.e., it indicates outflow (expansion) of the
upper layers of the extended atmosphere rather than a wind.

Along with the features common for the spectra of stars close to the AGB
stage, the H$\alpha$ profile in the spectrum of IRAS 20508 displays its own
individual peculiarities, first and foremost extended wings (Fig.\,2),
which are uncommon for such a cool supergiant. The H$\alpha$ emission is
superimposed with variable absorption, which is systematically shifted by
$V_r$=$V_r(H, abs)$-$V_r(met)\approx 10$\,km/s towards longer wavelengths
relative to the photospheric lines (only on the last date of our
observations does the position of the H absorption correspond to that of
the metallic lines). For all five observing epochs, the intensity of the
short-wavelength emission exceeds that of the long-wavelength emission.
Over five years of observations, the H$\alpha$ profile in the spectrum of
IRAS\,20508 varied from an intense, bell-shaped emission feature with only
a small amount of core absorption to a two-peak emission line.

It also follows from a comparison of the H profiles in Fig.\,2 that the
intensity of the emission systematically decreased during 1999--2003; the
central absorption line is below the continuum level in the 2003 spectrum.
The width of the emission wings of the line also varied. At our first four
epochs, when the emission flux decreased systematically relative to the
continuum radiation, an essentially constant difference between the radial
velocities derived from photospheric absorption lines and the center of
gravity of the H emission profile was observed:
$V_r=V_r(met)-V_r(H,emis) \approx -23$\,km/s. However,
the $V_r$(met) and $V_r$(H, emis)
velocities became virtually equal in the last spectrum obtained in 2004;
the H$\alpha$ emission intensity again increased, and the central absorption again
exceeded the continuum. Unfortunately, due to the absence of
radiospectroscopic observations of IRAS\,20508, we cannot determine the
systemic velocity of the source or suggest even a preliminary model for
the system. Variable H$\alpha$ emission is also a well-known phenomenon for AGB
and post-AGB stars [\cite{Oud94}] (see also references in the present study).
Differences in the H$\alpha$ profiles are due to differences in the dynamical
processes occurring in the extended atmospheres of these stars, such as
spherically symmetrical outflows with a velocity that is either constant
or variable with height in the atmosphere, matter falling onto the
photosphere, or pulsations. H-profile variability can be naturally
explained for a post-AGB star with signs of binarity and mass loss: the H$\alpha$
profile varies due to the orbital motion in the system. However, the H$\alpha$
profiles also vary in those post-AGB objects for which no signs of $V_r$
and brightness variability are detected. The brightness variability may
suggest a shock mechanism, like  RV\,Tauri stars, where the dissipation of
shocks in the atmosphere probably stimulates matter outflow. The radiative
fluxes from AGB and post-AGB stars are insufficient for the radiative
wind-generation mechanism, which is efficient in the case of hot massive
supergiants. The observed H profile can be considered a combination of two
lines with different natures: photospheric absorption and intense emission
with a broad set of velocities, forming in a circumstellar structure. In
this case, the emission component can be distinguished by subtracting the
theoretical photospheric profile from the observed profile. If we subtract
the photospheric profile calculated from a model with the solar H/He ratio
(indicated in Fig.\,3 by the dotted curve), the wings of the resulting
emission profile extend to radial velocities of $\pm$250\,km/s.

\subsubsection{Structure of the Na\,D profile.} A detailed study of
high-resolution spectra of IRAS\,20508 revealed that the Na\,D resonance
doublet also displays a complex profile, with both absorption and emission
components. In the spectra obtained with the PFES spectrograph, both the
Na\,D1 and Na\,D2 lines are unresolved asymmetric blends. Let us consider
the spectrum s40402 obtained with high spectral resolution $R\approx 60000$ with
the NES spectrograph. Fig.\,4 presents the observed spectrum in the area
of the Na D doublet, along with the theoretical spectrum and their
difference. A comparison of the observed and synthetic spectra shows
narrow absorption (indicated by 1 in Fig.\,4) and broad emission
(indicated by ``2'' in the difference spectrum in Fig.\,4) in the Na\,D
doublet. The position of the emission peak in the Na\,D doublet lines is
roughly consistent with the radial velocity estimated from the absorption
lines of metals. The half-width of the narrow absorption feature is a
factor of 1.5 smaller than the average half-width of the absorption lines
in the spectrum, suggesting that the narrow absorption feature is formed
in the interstellar medium. It follows from Table~2 that the velocity of
the narrow interstellar component is $V_r$=$-7.7$\,km/s (relative to the
Local Standard of Rest, $V_{lsr}$=9.3\,km/s). According to [\cite{Hobbs}],
interstellar Na\,D2 lines at positions corresponding to velocities $V_r
\approx -8$ . . . $-$20\,km/s are observed in the Galaxy, in the direction
towards the constellation Vulpecula, close to IRAS\,20508. Note that,
according to [\cite{Georg}], the radial velocities of HII regions in the
local volume of the Galaxy in the direction towards IRAS\,20508 are
$V_{lsr}$,HII 10\,km/s, which is close to our obtained value. Brand and
Blitz [\cite{Brand}] presented the velocity $V_{lsr},HII \approx 21$\,km/s
($V_r \approx 5$\,km/s) in this same direction, but for a more distant
region of the Galaxy with d=3.9\,kpc. 
The presence of a corresponding feature in the
spectrum of IRAS\,20508 cannot be excluded, in which case, this indicates
a larger distance to IRAS\,20508, as is also confirmed by the estimated
luminosity of the object (see Subsect.~3.2).

The presence of broad (half-widths reaching $V_r \approx 120$\,km/s) emission
components in the Na\,D lines suggests that IRAS\,20508 belongs to a
relatively small group of supergiants with this type of anomaly in their
spectra. Intense and broad (up to 300\,km/s) emission in resonance lines,
in particular Na\,D, is observed in the spectrum of R\,CrB [\cite{Rao}]
close to its minimum brightness, as well as in the spectrum of FG\,Sge
[\cite{Gonz, Kipper, Rao2002}], which is manifests itself as a R\,CrB-type
nova, as was shown by Gonzalez et al. [\cite{Gonz}]. The same spectral
peculiarity is also observed for a number of post-AGB stars, for example,
89\,Her [\cite{Rao2002}], QY\,Sge [\cite{Rao2002}], and V510\,Pup
[\cite{IRAS08005}]. In the case of the hotter photosphere of QY\,Sge, the
emission is very strong, far exceeding the continuum level
[\cite{Rao2002}]. Rao Kameswara et al. [\cite{Rao2002}] suggest that the
resonance-line emission is formed in hot circumstellar envelopes, while
the large width results from photon scattering on moving dust particles of
the envelope. The Na\,D emission in the spectrum of IRAS\,20508 is
distorted by intense absorption forming in the photosphere of the cool
supergiant. It is likely that the spectrum of IRAS\,20508 displays
absorption identified with diffuse interstellar bands (DIB). However, due
to the late spectral type of the star, even the strong 5780\,\AA{} band is
blended by photospheric lines, complicating distinguishing this band and
measuring its parameters. For the same reason, the radial velocity is
difficult to measure, even for the usually easily discriminated
6613\,\AA{} band. In particular, it is blended by the YII line in the
spectrum of the cool supergiant.

\subsection{Main Parameters of the Star}

To determine the basic parameters of the model atmosphere of the star--the
effective temperature Teff and gravitational acceleration log g--and
calculate the chemical composition and synthetic spectra, we used the grid
of model stellar atmospheres calculated by Kurucz [\cite{Kurucz}] in a
hydrostatic approximation for various metallicities. The effective
temperature was determined in the usual way, from the condition that the
abundance of neutral iron be independent of the excitation potentials for
the lines used. The gravitatonal acceleration was selected based on the
condition of ionization balance for the iron atoms, and the microturbulent
velocity from the condition that the iron abundance be independent of the
line intensities. A supplementary criterion for the reliability of the
method is that the same dependence as that for Fe not be derived for other
elements represented by numerous lines in the spectra (e.g., SiI, CaI,
TiI, CrI, NiI). In addition, if the microturbulent velocity has been
determined with high reliability, there should be no dependence between
the individual abundances and equivalent widths of the lines used for the
calculation. The abundances of titanium and vanadium derived from lines of
neutral atoms and ions are consistent within the uncertainties. This
provides evidence that the gravitational acceleration in the atmosphere
has been correctly estimated based on the condition that the iron atoms be
in ionization balance. Overall, the internal consistency of the parameters
suggests that the homogeneous model atmospheres used are adequate for
calculations of weak lines in the approximation of LTE. Table\,3 presents
the elemental abundances (X) for individual lines. Table\,4 contains the
resulting parameters of the model atmosphere $T_{eff}$, $\log g$, and
$\xi_t$ together with the average elemental abundances relative to iron
[X/Fe]. The abundances of elements in the solar photosphere are taken from
[\cite{Grevesse}]. The oscillator strengths of spectral lines involved in
determination of the model parameters and elemental abundances are given
in [\cite{AICMI, IRAS23304}]. The average typical uncertainties in the
model parameters for a star with an effective temperature near 5000\,K are
$T_{eff} \approx 100$\,K, $\log g \approx 0.3$\,dex, and $\xi_t \approx
0.5$\,km/s. The dispersion of the elemental abundances derived from a set
of lines is small: the rms deviation usually does not exceed 0.3\,dex
(Table 4). All our calculations were carried out using the WIDTH9 code
assuming LTE-approach. Corrections for superfine structure and isotopic
shifts, which broaden the NiI, MnI, and BaII lines, were not taken into
account. To verify the reliability of the derived parameters of the model
atmosphere, we compared the observed spectrum with the synthetic spectrum
calculated using the STARSP code [28]. The observed and theoretical
spectra (see example in Fig.\,1) are reasonably consistent. When
determining the model-atmosphere parameters, we used lines with low and
moderate intensity with equivalent widths $W \le 0.25$\,\AA{}, since the
approximation of a stationary plane-parallel atmosphere may be inadequate
to describe the strongest spectral features. In addition, some strong
absorption features may be distorted by the influence of the circumstellar
envelope, and if the spectral resolution is not sufficiently high, the
intensity of the envelope components will be included in the observed
intensities of components formed in the atmosphere. The effective
temperature of the central star $T_{eff}$=4800\,K indicates that
IRAS\,20508 is in the transition from the AGB towards the PPN stage [29].
The value $\log g$=1.5 testifies that the luminosity of the star is not
very high. We estimated the absolute magnitude $M_v$ from the equivalent
width of the OI 7773\,\AA{} oxygen IR triplet, $W$=0.57\,\AA. It is known
that the equivalent width of this triplet W(OI) may represent a good
indication of the absolute luminosities of supergiants over a broad
interval of temperatures. Using the calibration of Ferro et al. [30], we
obtain $M_v \approx -3^m$. Applying the calibration relations of Straizys
and Kuriliene [31], we can translate $T_{eff}$=4800\,K and $M_v \approx
-3^m$ into the star's spectral class and luminosity type: G5\,I--II, which
corresponds to the normal color index $(B-V)_o \approx 0.95$, which is
consistent with the observed value $B-V$=0.93. We thus conclude that no
color excess is observed for the high-latitude object IRAS\,20508. This
makes it possible to estimate the distance to the star from its absolute
luminosity: $d=V - M_v \approx $ 10\,kpc. Note here that, according to
Neckel and Klare [32], the interstellar absorption in the disk of the
Galaxy towards an object with longitude $l=66^o$ is substantial: even at
distances up to 1 kpc, the absorption increases to $A_v \ge 3^m$. The
absence of any substantial absorption at optical and even UV wavelengths,
despite the large amount of dust around the star, is one of the mysteries
of PPN. In this sense, the spectral energy distribution (SED) of the
supergiant HD\,161796=IRAS\,17436+5003, which displays a strong IR excess
[33] and no distortions of the SED at optical and UV wavelengths [34, 35]
is considered to be typical. In general, the absence of any substantial
absorption at UV wavelengths can be understood if the dust envelope is
spatially separated from the central star; thus, the envelope does not
substantially affect the radiation of the star. The large distance to
IRAS\,20508 and the absence of reddening with a large IR excess leads us
to con sider another aspect of the structure of supergiants with IR
excesses and their distances. Menzies and Whitelock [36] noted a
paradoxical difference between the distances to the G0I star QY Sge (the
optical component of IRAS 20056+1834) obtained using different techniques.
Menzies and Whitelock [36] obtained the distance $d \approx 500$\,pc for QY\,Sge based
on its interstellar absorption, while taking the absolute magnitude for a
population I G0I supergiant to be $M_v=-6\lefteqn{.}^m5$ yields a distance
to QY\,Sge of d$\approx$36\,kpc (for a low-mass population II supergiant
in the post--AGB stage, $d \approx 9$\,kpc). To explain this substantial difference,
Menzies and Whitelock [36] suggested a model with an extremely
inhomogeneous circumstellar envelope heated by the star. QY\,Sge is
considered to be a nearby system ($d \approx 500$\,pc), in which the direct radiation
from the star towards the observer is obscured by the dust envelope, while
the observed flux is determined by scattering on matter ejected by the
star earlier and located behind it. In such an aspherical system, the
observed flux essentially does not depend on the temperature and
luminosity of the star; instead, it is specified by the morphological
parameters of the system--the distance between the envelope and central
star, the angle between the system's axis and the line of sight, the angle
of the cone in which the star's radiation emerges, and the optical depth
and parameters of the dust particles of the envelope. Similar models were
suggested previously for several objects, such as the supergiant VY\,CMa
[37] and the AGB star CIT~3, which is identified with the IR source
IRC+1011 [38].

\subsection{Chemical abundances pattern}

Table~3 presents the measured equivalent widths W for individual lines in
one of the spectra (s25309), along with the calculated abundances (X) for
the entire set of lines used to determine the chemical composition. We
selected this spectrum (s25309) to calculate the chemical composition
because it has the highest signal-to-noise ratio (as can be seen in
Fig.\,1) and a large interval of detected wavelengths. Table~4 presents the
averaged abundances for 27 elements. Below, we will analyze in detail the
abundances of elements belonging to various groups; first, however, we
will consider the separation of elements in the atmosphere of an object
with a gas--dust envelope. Separation of elements in the envelope. The
studied object may be undergoing a stage of lowintensity mass exchange
between the atmosphere and circumstellar gas­dust envelope. However, the
very fact that the iron abundance is close to the solar value indicates
that condensation on dust particles does not cause any serious distortions
in the elemental abundances, since iron is among those elements that most
efficiently condense onto particles [39]. The CNO-group elements sulfur
and zinc are essentially unaffected by fractioning processes. Since the
abundances of CNO elements may vary due to nuclear reactions in the course
of the star's evolution, the behavior of zinc and sulphur is critical for
the efficiency of selective separation. The abundance of zinc does not
vary in the course of stellar nucleosynthesis in the interiors of low- and
medium-mass stars. In combination with the ratio $[Zn/Fe] = -0.24$, which
is not appreciably different from its solar value, this suggests that we do
not see any selective separation in the circumstellar envelope of
IRAS\,20508. The abundance of sulfur in the atmosphere of IRAS\,20508 is
very high: $[S/Fe] = +0.74$. A similar substantial sulfur excess was
detected, for example, in the stars HD\,331319 = IRAS\,19475+3119 and
HD\,161796 = IRAS\,17436+5003 [40], as well as the optical component of
the IR source IRAS\,04296+3429 [41]. A similar result for sulfur was also
obtained for the two evolved stars K\,413 [42] and K\,307 (a W\,Vir-type
$V1$ variable) [43], which are members of the globular cluster M\,12. Bond
and Luck [44] revealed a very large sulfur excess ($[S/Fe] = +1.2$) in the
atmosphere of the lowmetallicity post-AGB star HD\,46703, and suggested
that sulfur was synthesized via the capture of particles by $^{12}C$
nuclei. However, Klochkova [45] also detected an excessive abundance of
sulfur in the atmosphere of the normal massive supergiant $\alpha$~Per,
which does not have a circumstellar envelope, so that no condensation
should be present, while we do not expect the synthesis of sulfur at such
an early evolutionary stage. There are grounds to conclude that the sulfur
excess is a stable peculiarity in the chemical compositions of evolved
stars. Note that this excess is observed in stars with different
metallicities and temperatures. Light elements. Let us now consider the
abundances of light metals. No LiI 6707\,\AA{} line is seen in the
spectrum of IRAS\,20508. The abundances of CNO-group elements in its
atmosphere are substantially different from the solar values, though these
results have only low accuracy. The excess abundance of carbon $[C/Fe]
+0.9$ is determined from four weak CI lines with equivalent widths $W >
10$\,m\AA{}. In the case of nitrogen, only the NI\,7468\,\AA{} line was
measured, since the NI\,7423\,\AA{} line is blended with the SiI line and NI
7442\,\AA{} line is beyond the limits of our echelle frame. Therefore, the
derived excess of nitrogen should not be considered firm. We calculated
the oxygen abundance ([O/Fe] = +1.79) using reliably measured lines of the
OI\,7773\,\AA{} oxygen IR triplet. However, it is known that the lines of the
7773\,\AA{} triplet are sensitive to deviations from LTE. The 7773\,\AA{} lines
in the spectrum of IRAS\,20508 are not very strong, and their total
equivalent width is $W = 0.56$\,\AA{}. Taking into account the calculations of
Gratton et al. [46], we can neglect systematic errors in the oxygen
abundance derived from the triplet lines assuming LTE for such a cool
star, and consider the derived excess to be real. As a result, we obtain
$[C/O] = -0.9$ which indicates that IRAS 20508 is a member of the group of
evolved objects with oxygenenriched atmospheres. This result is consistent
with the position of the object in the $VI\,b$ region in the van der
Veen--Habing diagram [47]. Of the even -process elements, we determined
the abundances of Mg, Si, Ca, and Ti, in addition to sulfur (see above).
The Mg, Ca, and Ti abundances are slightly lower in comparison with the
iron abundance; the average excess for these elements is $[X/Fe] = -0.18$.
Sodium is overabundant in the atmosphere of IRAS\,20508, $[Na/Fe] = +0.58$,
which may be due to the synthesis of sodium in the course of hydrogen
burning [48, 49]. Some of the derived Na excess is probably due to our
failure to take into account superionization of the sodium atoms [50]. The
small derived aluminium excess, $[Al/Fe] = +0.32$ is inconsistent with the
oxygen excess, as was previously obtained for K\,413 [42]: an Al--O
anticorrelation [51] is usually observed for evolved stars in globular
clusters with average deficits of metals. Iron peak elements. The average
abundances of iron-group metals (vanadium, chrom, manganese, cobalt,
nickel) differ little from the iron abundance: $[met/Fe] = -0.04$.
Individual iron-group elements display no significant deviations towards
either over- or underabundances relative to iron. The abundance of copper
is slightly low for both stars ($[Cu/Fe] = -0.23$); this value is determined
from only two lines, and its significance is fairly low. Heavy metals. It
is generally accepted (see, for example, [52]) that the atmospheres of AGB
and post-AGB stars may display excesses of heavy metals, due to neutron
capture, mixing, and the ejection of matter that has undergone these
processes from the stellar interior to its surface. However, in contrast
to the expected excess, the atmosphere of the optical component of IRAS
20508 displays a reduced abundance of barium: $[Ba/Fe] \approx -0.4$ (Tables 3, 4).
The abundances of the lighter $s$-process metals Y and Zr are also reduced
relative to that of iron: $[s/Fe] = -0.28$. The deficit is less severe for
the lanthanides La, Ce, Pr, Nd: $[lant/Fe] = -0.11$. A similar weak deficit
was obtained for the r-process element Eu: $[Eu/Fe] = -0.07$. Deficits of
$s$-process elements are observed in the atmospheres of post-AGB stars much
more frequently than excesses [1, 14, 45, 53]. It is most likely that the
observed absence of any manifestation of the ejection of heavy metals is
real, rather than being due to systematic errors in analyses of the
spectra of supergiants based on model atmospheres. The presence or absence
of an excess of s-process elements is probably related to the main parame
ters of the star, namely, its initial mass and massloss rate, which
specify its evolution. Gonzalez and Wallerstein [54] suggest that the
basic parameter determining the efficiency of the ejection of the products
of nucleosynthesis to surface layers is a star's luminosity. We have
obtained further support for this suggestion, since the luminosity of
IRAS\,20508 is not very high: $M_v \approx -3^m$.

\subsection{Evolutionary status of IRAS 20508}

As we noted earlier, according to its the position in the IR
color diagram, IRAS\,20508 belongs to the $VI\,b$ group in the classification
of van der Veen and Habing [47]. This group includes evolved stars, which
are often variable and have oxygen-enriched circumstellar envelopes. AGB
objects are represented by Mira-type stars, carbon stars, and OH/IR stars.
The Miras possess comparatively hot envelopes, are associated with $H_2O$
masers and maser sources radiating mainly in the main $OH$ band, and descend
from stars that have undergone mass loss at rates below $10^{-5}M_{\odot}/yr$ [55].
OH/IR stars are thought to be the final stage of evolution of oxygenrich
AGB stars that have undergone mass loss at larger rates (exceeding
$10^{-5} M_{\odot}/yr $) and are rapidly evolving towards the PPN stage. Unfortunately, no
observations in water maser or $OH$ bands are available for IRAS\,20508,
hindering unambiguous classification of the envelope. However, it may be
that, in the case of IRAS\,20508, we are observing an extremely early phase
of PPN formation, immediately after the termination of mass loss and the
beginning of the separation of the envelope. Excesses of neither lithium
nor $s$-process elements were detected in the atmosphere of IRAS 20508. The
mixing that took place in the object is only indicated by the
overabundance of $CNO$--group elements, which are products of helium burning.
In the case of IRAS\,20508, we have obtained further confirmation of the
well known corellation [14, 56] between the $C/O$ ratio and the abundance of
s-process elements in the atmospheres of and post-AGB stars. Note also
that the parameters of the atmosphere of IRAS\,20508, its metallicity, and
details of its chemical composition essentially coincide with those for
the OH/IR star associated with IRAS\,18123+0511 [57].

\section{Conclusions}

Our high-resolution optical spectroscopic observations made with the 6-m
SAO telescope have revealed spectral variability of the cool star with a
high absolute luminosity associated with the IR source IRAS\,20508+2011. We
have identified lines in the spectrum at wavelengths from 4300 to 7930\,\AA{},
and measured the equivalent widths and radial velocities of numerous
absorption lines of neutral atoms and ions. The radial velocity of the star
derived from the photospheric absorption lines is variable: over five
years of observations, it varied in the interval 15--30 km/s. Over the
same time, the H$\alpha$ absorption-emission profile also varied substantially: a
bell-shaped emission line with only a small amount of core absorption was
transformed into a two-peak emission feature with central absorption below
the continuum. Further observations with higher spectral resolution are
required to explain this variability. The Na\,D doublet lines also display
a complex profile that consists of broad (with half-width 120\,km/s)
emission, whose position corresponds to that of photospheric absorption
lines in the spectrum, and narrow absorption with an interstellar origin.
The derived parameters of the star (its luminosity, effective temperature
$T_{eff} = 4800$\,K, gravitational acceleration $\log g = 1.5$, microturbulent
velocity $\xi_t = 4.0$\,km/s, and metallicity $[Fe/H] = -0.36$) indicate that the
optical component of IRAS\,20508+2011 is a star that is close to the AGB
stage with an absolute magnitude of $M_v -3^m$ at a distance of $d \approx 10$\,kpc.

The chemical composition of its atmosphere is fairly usual for this
evolutionary stage. The relation between the carbon excess $[C/Fe] \approx +0.9$ and
oxygen excess $[O/Fe] = +1.79$ suggests that IRAS\,20508 is a member of the
group of evolved stars with oxygen-enriched atmospheres ($[C/O] = -0.9$),
consistent with its position in the IR colour diagram. No excess of
$s$-process elements is observed, which correlates well with the low
absolute luminosity of the star and the ratio of its carbon and oxygen
abundances, [C/O].

\section{Acknowledgements}

This work was supported by the Russian Foundation for Basic Research
(project No.\,03--02--39019), the Basic Research Program of the Department
of Physical Sciences of the Russian Academy of Sciences ``Extended Objects
in the Universe'' (``Spectroscopy of Extended Envelopes of Stars in Late
Stages of Their Evolution''), and the grant from the National Science
Foundation of China (No.\,10433010). This publication is based on work
supported by Award No.\,RUP1--2687--NA--05 of the U.S. Civilian Research \&
Development Foundation (CRDF).
This research has used the SIMBAD
database and ALADIN interactive sky atlas of the Strasbourg Center for
Astronomical Data (CDS).

\newpage

\begin{figure}[t]	      		      
\includegraphics[angle=-90,width=1.0\textwidth,bb=37 150 560 800,clip]{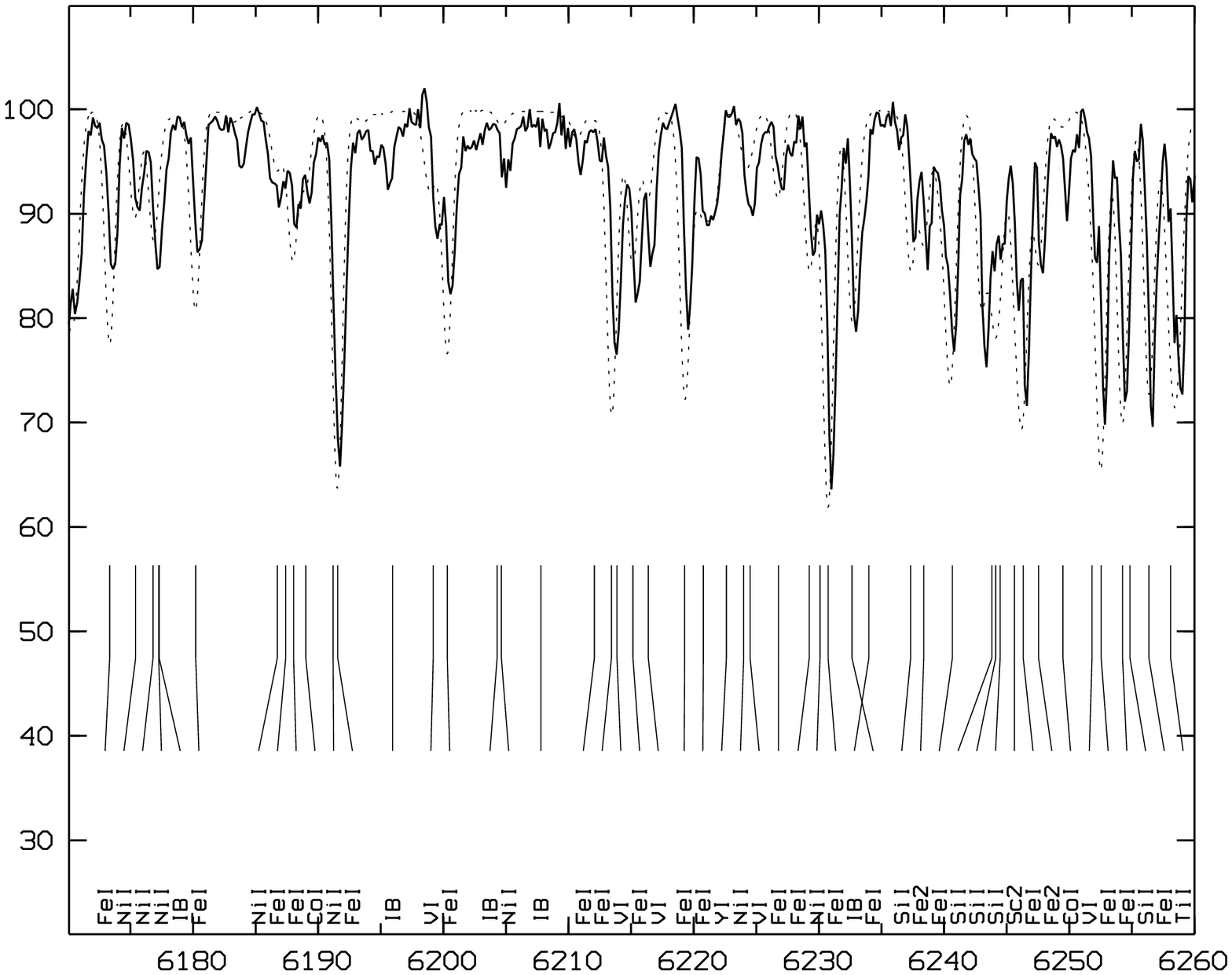}
\caption{Fragment of the observed spectrum of IRAS\,20508+2011 (solid) and
       the theoretical spectrum (dashed) calculated with the model parameters
       $T_{eff}$=4800\,K, $\log g$=1.5, $\xi_t$=4.0\,km/s.}
\label{6170}
\end{figure}

\begin{figure}[t]	      		      
\includegraphics[angle=0,width=0.8\textwidth,bb=90 70 555 780,clip]{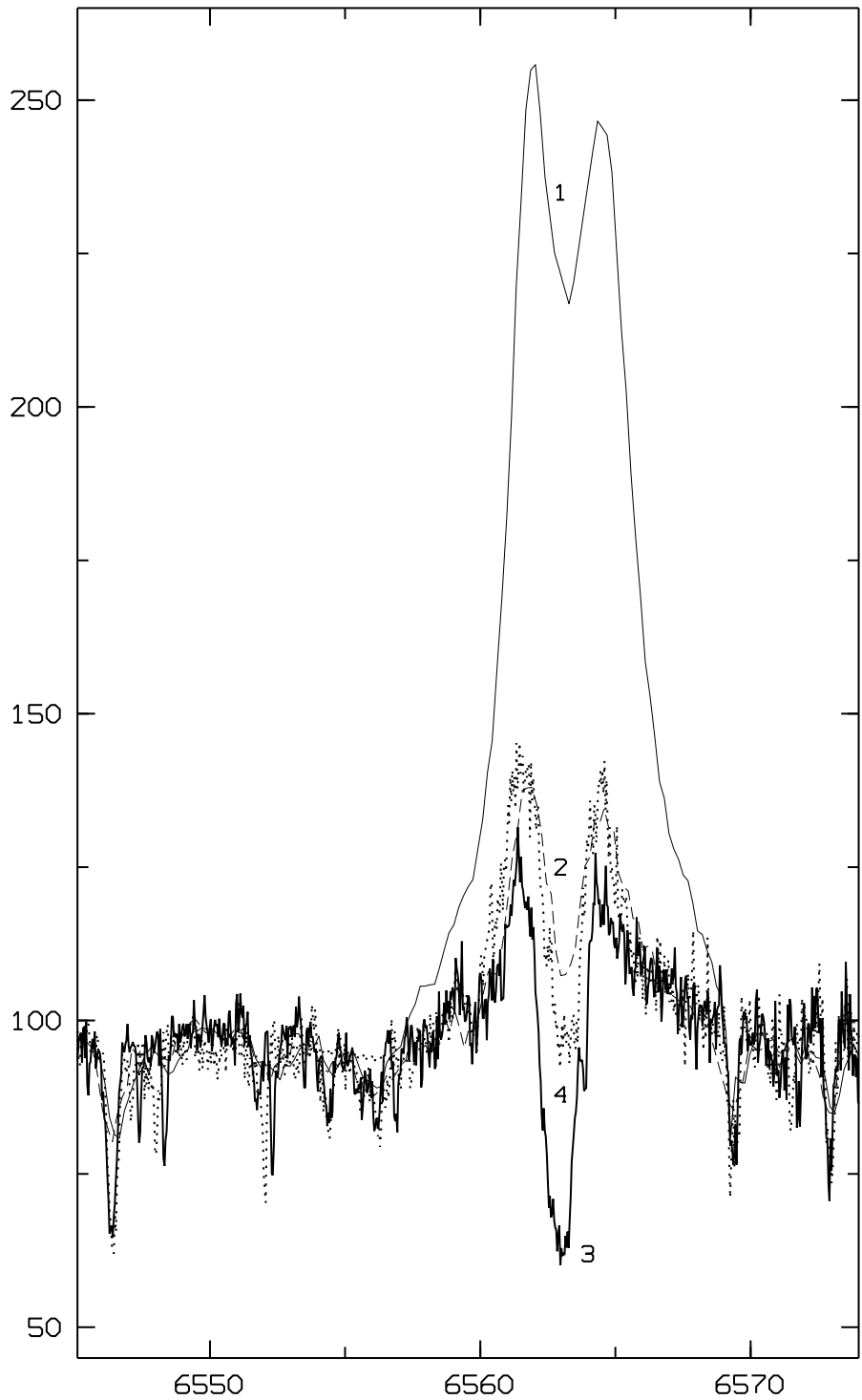}
\caption{H$\alpha$ profile in spectra of IRAS\,20508+2011 obtained in (1) 1999,
       (2) 2000, (3) 2003, and (4) 2004.}
\label{Halpha}
\end{figure}

\clearpage
\newpage

\begin{figure}	      		      
\includegraphics[angle=-90,width=1.0\textwidth,bb=40 150 540 790,clip]{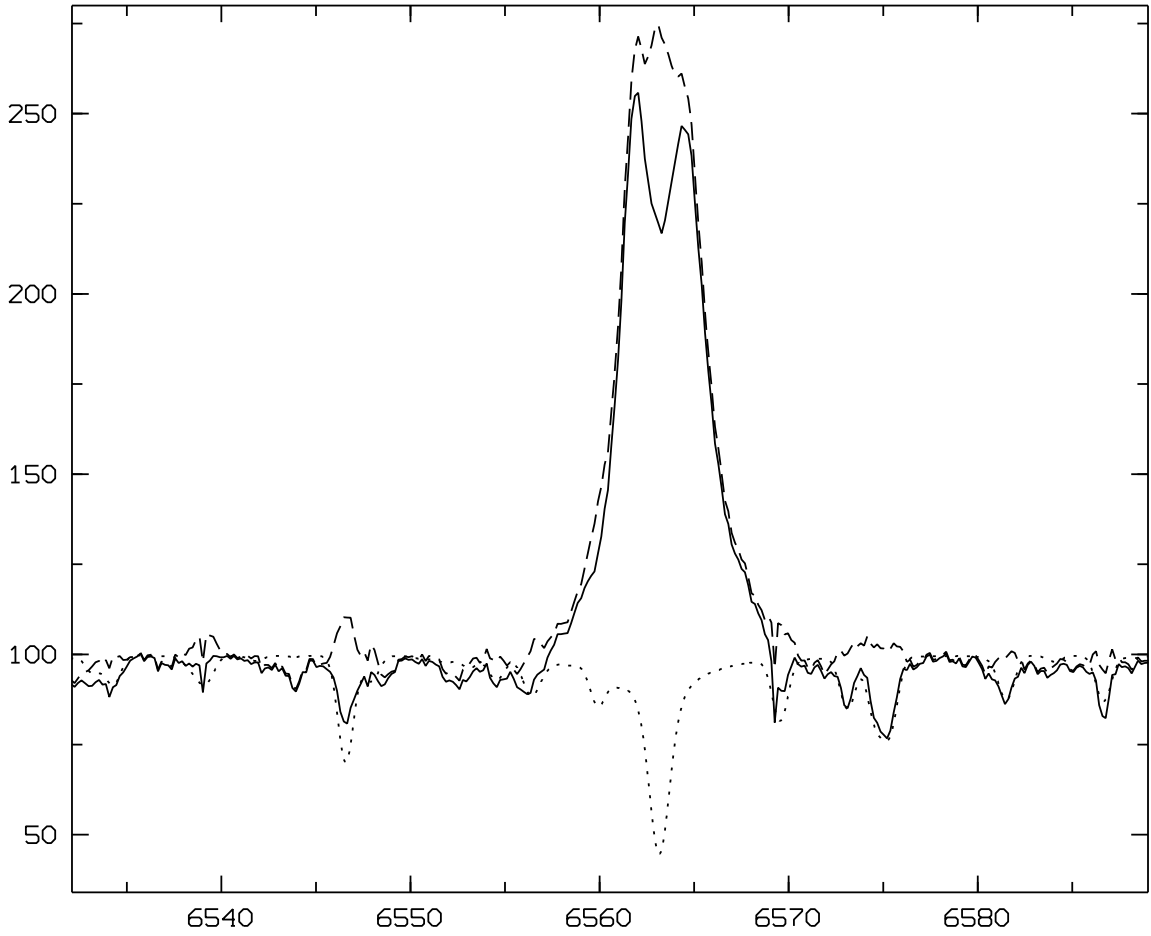}
\caption{H$\alpha$ line profile in spectrum s25309 of IRAS\,20508+2011 (solid)
together with the theoretical spectrum (dotted). The dashed curve
indicates the difference between the observed and theoretical spectra.}
\label{alpha}
\end{figure}

\begin{figure}	      		      
\includegraphics[angle=-90,width=1.0\textwidth,bb=50 150 560 800,clip]{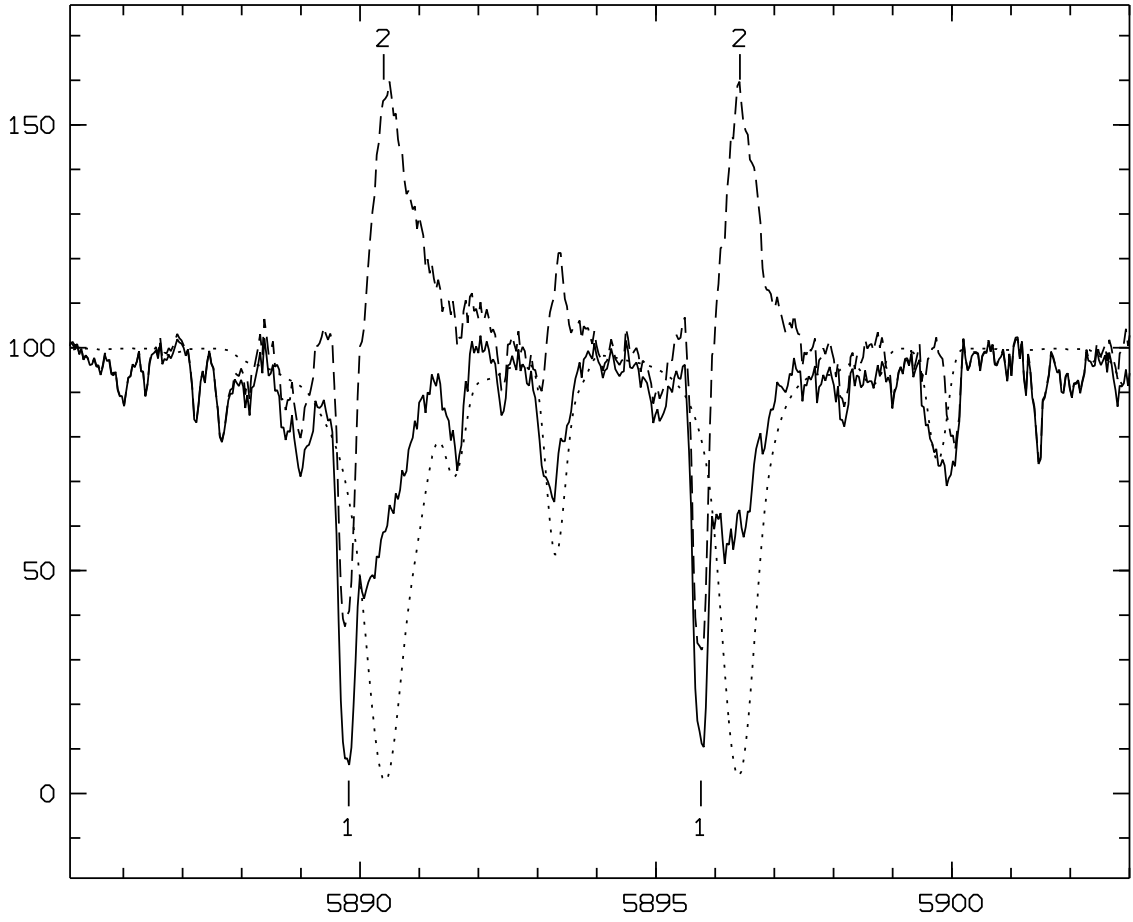}
\caption{Same as Fig.\,3 for the Na\,D line in spectrum s40402.
        Figures ``1'' and ``2'' indicate the positions of narrow absorption and
        broad emission features (see the text).}
\label{Naem}
\end{figure}

\clearpage
\newpage

\begin{table}
\centering
\caption{Observations of IRAS\,20508+2011 with the 6\,m telescope}
\medskip
\begin{tabular}{l@{\quad}c@{\quad}c@{\quad}c@{\quad}l}
\hline
Spectrum & Date     &  UT   & $\Delta\lambda$, \AA{} & Spectrograph\\
\hline
s25309 &   28.07.1999 & 21:50 & 4550--7930  & PFES \\
s28208 &   04.07.2000 & 23:13 & 4300--7820  & PFES \\
s28412 &   06.07.2000 & 23:35 & 4320--7825  & PFES \\
s40402 &   10.08.2003 & 23:05 & 5270--6760  & NES  \\
s43223 &   27.08.2004 & 22:51 & 5270--6760  & NES  \\
\hline
\end{tabular}
\label{obs}
\end{table}

\clearpage
\newpage

\begin{table}[h]
\caption{Mean heliocentric radial velocities for line groups and some individual lines
        derived from IRAS\,20508+2011 spectra obtained on various dates.
	 Uncertain values are given by Italic.}
\bigskip
\begin{tabular}{l@{\quad}r@{\quad}r@{\quad}r@{\quad}r@{\quad}r}
\hline
Spectral & \multicolumn{4}{c}{${\rm Vr_{\odot}}$, km/s }     \\
\cline{2-6} 
features     & 28.07.99 &04.07.00 & 06.07.00 & 10.08.03. &27.08.04\\
\hline
\multicolumn{4}{l}{\underline{Absorptions}} &           &        \\
metals     & 15.5     &29.8     & 27.1     & 20.3      & 29.9   \\
H$\alpha$  & 23.1     &38.6     & 40.7     & 21.4      & 38.0   \\
H$\beta$   & 16.5     &27.0     & 24.4     &           &        \\
H$\gamma$  &          &27.6     & 23.7     &           &        \\
\hline  						        
\multicolumn{4}{l}{\underline{Emission}} &  &  \\  	        
 H$\alpha$ &39.3      & 51.6    &{\it 57.1}& 43.1      & 27.0   \\
\hline							        
I.S. NaI(D1,D2)&$-4.6$&$ -5.9$  & $-5.5$   &$-7.7$     &$-7.3$   \\ 
\hline                                                           
\end{tabular}
\label{RV}
\end{table}

\clearpage
\newpage

\begin{center}
\tablecaption{Lines identification  and their intensities in spectrum s25309 of IRAS\,20508+2011
               and the  elemental abundances derived}
\tablehead{\hline $\lambda$, \AA{} & Ion & EP, eV &   gf   & W$_{\lambda}$,m\AA{} & $\epsilon$(X)  \\
 [2pt] \hline }
\tabletail{\hline \rule{0pt}{5pt}&&&&& \\} 
\begin{supertabular}{c lr@{\quad}  @{\quad} r @{\quad}r@{\quad}l@{\quad} }
 6014.84 &   CI   &   8.64 & $ -1.71$ &   13  &  $ -3.00$   \\  
 6671.82 &   CI   &   8.85 & $ -1.65$ &   16  &  $ -2.69$   \\  
 7100.30 &   CI   &   8.64 & $ -1.60$ &   19  &  $ -2.81$   \\  
 7115.19 &   CI   &   8.64 & $ -0.90$ &   29  &  $ -3.23$   \\  
 7468.31 &   NI   &  10.34 & $ -0.13$ &   27  &  $ -2.67$   \\  
 7771.94 &   OI   &   9.15 & $  0.33$ &  210  &  $ -1.73$   \\  
 7774.17 &   OI   &   9.15 & $  0.19$ &  214  &  $ -1.48$   \\  
 7775.39 &   OI   &   9.15 & $ -0.03$ &  147  &  $ -1.91$   \\  
 4668.57 &   NaI  &   2.10 & $ -1.25$ &  161  &  $ -5.40$   \\  
 4751.83 &   NaI  &   2.10 & $ -2.10$ &   91  &  $ -5.19$   \\  
 5682.65 &   NaI  &   2.10 & $ -0.71$ &  229  &  $ -5.42$   \\  
 4703.00 &   MgI  &   4.34 & $ -0.38$ &  268  &  $ -5.20$   \\  
 5711.10 &   MgI  &   4.34 & $ -1.68$ &  136  &  $ -5.28$   \\  
 7657.61 &   MgI  &   5.11 & $ -1.28$ &  134  &  $ -4.90$   \\  
 6696.03 &   AlI  &   3.14 & $ -1.32$ &  136  &  $ -5.48$   \\  
 7835.32 &   AlI  &   4.02 & $ -0.47$ &  113  &  $ -5.53$   \\  
 7836.13 &   AlI  &   4.02 & $ -0.29$ &  116  &  $ -5.69$   \\  
 5665.56 &   SiI  &   4.92 & $ -2.04$ &   42  &  $ -5.06$   \\  
 5684.49 &   SiI  &   4.95 & $ -1.65$ &  142  &  $ -4.46$   \\  
 5690.43 &   SiI  &   4.93 & $ -1.87$ &   64  &  $ -4.96$   \\  
 5772.15 &   SiI  &   5.08 & $ -1.75$ &  100  &  $ -4.58$   \\  
 5793.08 &   SiI  &   4.93 & $ -2.06$ &   83  &  $ -4.59$   \\  
 5948.55 &   SiI  &   5.08 & $ -1.23$ &  160  &  $ -4.57$   \\  
 6087.79 &   SiI  &   5.87 & $ -1.71$ &   20  &  $ -4.72$   \\  
 6125.03 &   SiI  &   5.61 & $ -1.51$ &   55  &  $ -4.67$   \\  
 6145.02 &   SiI  &   5.61 & $ -1.48$ &   43  &  $ -4.84$   \\  
 6237.33 &   SiI  &   5.61 & $ -1.14$ &   92   & $  -4.6$7  \\  
 6721.84 &   SiI  &   5.56 & $ -1.26$ &   63   & $  -4.9$0  \\  
 7034.91 &   SiI  &   5.87 & $ -0.88$ &   61   & $  -4.9$6  \\  
 7405.79 &   SiI  &   5.61 & $ -0.54$ &  150   & $  -4.8$0  \\  
 7800.00 &   SiI  &   6.18 & $ -0.71$ &  101   & $  -4.4$2  \\  
 6347.09 &   SiI  &   8.12 & $  0.26$ &   61   & $ -4.48$   \\  
 5696.63 &   SI   &   7.87 & $ -1.21$ &   16   & $ -4.15$   \\  
 6045.99 &   SI   &   7.87 & $ -0.79$ &   18   & $ -4.47$   \\  
 6052.63 &   SI   &   7.87 & $ -0.63$ &   16   & $ -4.69$   \\  
 6743.58 &   SI   &   7.87 & $ -0.70$ &   43   & $ -3.93$   \\  
 6757.16 &   SI   &   7.87 & $ -0.29$ &   22   & $ -4.81$   \\  
 5261.71 &   CaI  &   2.52 & $ -0.58$ &  174   & $ -6.30$   \\  
 5349.47 &   CaI  &   2.71 & $ -0.31$ &  214   & $ -5.96$   \\  
 5581.98 &   CaI  &   2.52 & $ -0.56$ &  223   & $ -5.88$   \\  
 5588.76 &   CaI  &   2.52 & $  0.36$ &  284   & $ -6.19$   \\  
 5590.13 &   CaI  &   2.52 & $ -0.57$ &  232   & $ -5.78$   \\  
 5594.47 &   CaI  &   2.52 & $  0.10$ &  291   & $ -5.87$   \\  
 5598.49 &   CaI  &   2.52 & $ -0.09$ &  277   & $ -5.82$   \\  
 5601.29 &   CaI  &   2.52 & $ -0.52$ &  218   & $ -5.98$   \\  
 5867.57 &   CaI  &   2.93 & $ -1.79$ &   67   & $ -5.60$   \\  
 6162.18 &   CaI  &   1.90 & $ -0.09$ &  369   & $ -6.01$   \\  
 6166.44 &   CaI  &   2.52 & $ -1.14$ &  114   & $ -6.35$   \\  
 6439.08 &   CaI  &   2.52 & $  0.39$ &  321   & $ -6.09$   \\  
 6471.67 &   CaI  &   2.52 & $ -0.69$ &  219   & $ -5.93$   \\  
 6499.65 &   CaI  &   2.52 & $ -0.82$ &  208   & $ -5.89$   \\  
 6508.84 &   CaI  &   2.53 & $ -2.65$ &   27   & $ -5.75$   \\  
 6572.80 &   CaI  &   0.00 & $ -4.31$ &  145   & $ -6.07$   \\  
 6717.69 &   CaI  &   2.71 & $ -0.52$ &  251   & $ -5.60$   \\  
 7202.21 &   CaI  &   2.71 & $ -0.26$ &  212   & $ -6.25$   \\  
 5239.81 &   ScII &   1.45 & $ -0.77$ &  129   & $ -9.60$   \\  
 5357.19 &   ScII &   1.51 & $ -2.21$ &   23   & $ -9.27$   \\  
 5526.79 &   ScII &   1.77 & $  0.13$ &  170   & $ -9.77$   \\  
 5669.04 &   ScII &   1.50 & $ -1.09$ &  109   & $ -9.43$   \\  
 6245.62 &   ScII &   1.51 & $ -0.93$ &  151   & $ -9.26$   \\  
 6320.84 &   ScII &   1.50 & $ -1.71$ &   56   & $ -9.34$   \\  
 6604.58 &   ScII &   1.36 & $ -1.48$ &  120   & $ -9.18$   \\  
 5009.65 &   TiI  &   0.02 & $ -2.20$ &  147   & $ -7.48$   \\  
 5020.03 &   TiI  &   0.84 & $ -0.36$ &  190   & $ -7.90$   \\  
 5038.40 &   TiI  &   1.43 & $  0.07$ &  197   & $ -7.50$   \\  
 5043.58 &   TiI  &   0.84 & $ -1.68$ &  117   & $ -7.21$   \\  
 5147.48 &   TiI  &   0.00 & $ -1.96$ &  157   & $ -7.69$   \\  
 5194.04 &   TiI  &   2.10 & $ -0.50$ &   52   & $ -7.44$   \\  
 5201.10 &   TiI  &   2.09 & $ -0.69$ &   42   & $ -7.38$   \\  
 5338.33 &   TiI  &   0.83 & $ -1.81$ &   44   & $ -7.77$   \\  
 5426.26 &   TiI  &   0.02 & $ -2.95$ &   44   & $ -7.64$   \\  
 5460.50 &   TiI  &   0.05 & $ -2.75$ &   85   & $ -7.42$   \\  
 5490.15 &   TiI  &   1.46 & $ -0.88$ &  107   & $ -7.37$   \\  
 5514.54 &   TiI  &   1.44 & $ -0.50$ &  100   & $ -7.83$   \\  
 5644.14 &   TiI  &   2.27 & $  0.00$ &  103   & $ -7.31$   \\  
 5689.47 &   TiI  &   2.30 & $ -0.41$ &   50   & $ -7.36$   \\  
 5716.45 &   TiI  &   2.30 & $ -0.64$ &   16   & $ -7.70$   \\  
 5739.46 &   TiI  &   2.25 & $ -0.54$ &   29   & $ -7.57$   \\  
 5739.98 &   TiI  &   2.24 & $ -0.61$ &   54   & $ -7.19$   \\  
 5766.33 &   TiI  &   3.29 & $  0.31$ &   29   & $ -7.20$   \\  
 5774.04 &   TiI  &   3.31 & $  0.54$ &   26   & $ -7.46$   \\  
 5839.76 &   TiI  &   1.46 & $ -2.36$ &   06   & $ -7.43$   \\  
 5866.45 &   TiI  &   1.07 & $ -0.78$ &  117   & $ -7.92$   \\  
 5903.32 &   TiI  &   1.07 & $ -2.09$ &   46   & $ -7.22$   \\  
 5922.11 &   TiI  &   1.05 & $ -1.41$ &  115   & $ -7.33$   \\  
 5937.81 &   TiI  &   1.07 & $ -1.83$ &   65   & $ -7.29$   \\  
 5944.68 &   TiI  &   0.00 & $ -3.79$ &   14   & $ -7.43$   \\  
 5965.83 &   TiI  &   1.88 & $ -0.35$ &  100   & $ -7.49$   \\  
 5978.54 &   TiI  &   1.87 & $ -0.44$ &   92   & $ -7.47$   \\  
 5988.56 &   TiI  &   1.89 & $ -1.12$ &   46   & $ -7.20$   \\  
 6064.63 &   TiI  &   1.05 & $ -1.89$ &   65   & $ -7.26$   \\  
 6126.22 &   TiI  &   1.07 & $ -1.37$ &   68   & $ -7.73$   \\  
 6258.10 &   TiI  &   1.44 & $ -0.30$ &  133   & $ -7.85$   \\  
 6261.10 &   TiI  &   1.43 & $ -0.42$ &  147   & $ -7.64$   \\  
 6303.75 &   TiI  &   1.44 & $ -1.51$ &   42   & $ -7.41$   \\  
 6743.12 &   TiI  &   0.90 & $ -1.57$ &   87   & $ -7.63$   \\  
 7138.91 &   TiI  &   1.44 & $ -1.53$ &   51   & $ -7.35$   \\  
 4708.65 &   TiII &   1.24 & $ -2.21$ &  137   & $ -7.90$   \\  
 4779.98 &   TiII &   2.05 & $ -1.37$ &  152   & $ -7.67$   \\  
 5185.90 &   TiII &   1.89 & $ -1.35$ &  149   & $ -7.92$   \\  
 5211.53 &   TiII &   2.59 & $ -1.85$ &   88   & $ -7.21$   \\  
 5336.78 &   TiII &   1.58 & $ -1.70$ &  172   & $ -7.73$   \\  
 5381.01 &   TiII &   1.57 & $ -2.08$ &  161   & $ -7.50$   \\  
 5418.80 &   TiII &   1.58 & $ -2.17$ &  131   & $ -7.67$   \\  
 6606.97 &   TiII &   2.06 & $ -2.79$ &   44   & $ -7.38$   \\  
 5670.85 &   VI   &   1.08 & $ -0.42$ &   42   & $ -8.91$   \\  
 5703.59 &   VI   &   1.05 & $ -0.21$ &   72   & $ -8.86$   \\  
 5727.05 &   VI   &   1.08 & $ -0.01$ &  126   & $ -8.60$   \\  
 5727.65 &   VI   &   1.05 & $ -0.87$ &   79   & $ -8.14$   \\  
 5737.06  &  VI   &   1.06 & $ -0.74$ &   73   & $ -8.30$   \\  
 5830.68  &  VI   &   3.11 & $  0.62$ &   10   & $ -8.23$   \\  
 6039.73  &  VI   &   1.06 & $ -0.65$ &   78   & $ -8.37$   \\  
 6090.21  &  VI   &   1.08 & $ -0.06$ &   98   & $ -8.79$   \\  
 6119.53  &  VI   &   1.06 & $ -0.32$ &   69   & $ -8.79$   \\  
 6150.16  &  VI   &   0.30 & $ -1.78$ &   65   & $ -8.31$   \\  
 6199.19  &  VI   &   0.29 & $ -1.30$ &  100   & $ -8.54$   \\  
 6216.37  &  VI   &   0.28 & $ -1.29$ &  130   & $ -8.36$   \\  
 6251.82  &  VI   &   0.29 & $ -1.34$ &   94   & $ -8.55$   \\  
 6266.32  &  VI   &   0.28 & $ -2.29$ &   40   & $ -8.12$   \\  
 6274.65  &  VI   &   0.27 & $ -1.67$ &   35   & $ -8.81$   \\  
 6285.16  &  VI   &   0.28 & $ -1.51$ &   48   & $ -8.80$   \\  
 6504.16  &  VI   &   1.18 & $ -1.23$ &   39   & $ -8.08$   \\  
 5819.93  &  VII  &   2.52 & $ -1.70$ &   18   & $ -8.50$   \\  
 6028.27  &  VII  &   2.49 & $ -1.98$ &   22   & $ -8.16$   \\  
 6029.00  &  VII  &   2.56 & $ -1.94$ &   09   & $ -8.55$   \\  
 5214.14  &  CrI  &   3.37 & $ -0.74$ &   38   & $ -6.68$   \\  
 5221.76  &  CrI  &   3.38 & $ -0.57$ &   51   & $ -6.68$   \\  
 5238.97  &  CrI  &   2.71 & $ -1.30$ &   77   & $ -6.48$   \\  
 5272.01  &  CrI  &   3.45 & $ -0.42$ &   71   & $ -6.55$   \\  
 5300.74  &  CrI  &   0.98 & $ -2.12$ &  240   & $ -6.34$   \\  
 5304.19  &  CrI  &   3.46 & $ -0.69$ &   45   & $ -6.54$   \\  
 5312.88  &  CrI  &   3.45 & $ -0.56$ &   82   & $ -6.31$   \\  
 5628.64  &  CrI  &   3.42 & $ -0.77$ &   58   & $ -6.38$   \\  
 5783.11  &  CrI  &   3.32 & $ -0.50$ &  101   & $ -6.38$   \\  
 5783.89  &  CrI  &   3.32 & $ -0.29$ &  116   & $ -6.47$   \\  
 6501.20  &  CrI  &   0.98 & $ -3.66$ &   56   & $ -6.49$   \\  
 6630.00  &  CrI  &   1.03 & $ -3.56$ &   32   & $ -6.84$   \\  
 6680.15  &  CrI  &   4.16 & $ -0.52$ &   32   & $ -6.15$   \\  
 6789.15  &  CrI  &   3.84 & $ -1.17$ &   14   & $ -6.27$   \\  
 6925.22  &  CrI  &   3.45 & $ -0.33$ &   92   & $ -6.54$   \\  
 6979.81  &  CrI  &   3.46 & $ -0.41$ &  114   & $ -6.27$   \\  
 5237.35  &  CrII &   4.07 & $ -1.16$ &   89   & $ -7.07$   \\  
 5308.46  &  CrII &   4.07 & $ -1.81$ &  101   & $ -6.29$   \\  
 5310.73  &  CrII &   4.07 & $ -2.28$ &   56   & $ -6.32$   \\  
 5313.61  &  CrII &   4.07 & $ -1.65$ &   63   & $ -6.87$   \\  
 5334.88  &  CrII &   4.07 & $ -1.89$ &   63   & $ -6.62$   \\  
 5508.63  &  CrII &   4.16 & $ -2.11$ &   59   & $ -6.36$   \\  
 5394.68  &  MnI  &   0.00 & $ -3.50$ &  179   & $ -7.59$   \\  
 5399.47  &  MnI  &   3.85 & $ -0.64$ &   20   & $ -7.28$   \\  
 5407.43  &  MnI  &   2.14 & $ -1.74$ &  161   & $ -6.79$   \\  
 5420.37  &  MnI  &   2.14 & $ -1.46$ &  186   & $ -6.84$   \\  
 5432.56  &  MnI  &   0.00 & $ -3.80$ &  150   & $ -7.52$   \\  
 5457.47  &  MnI  &   2.16 & $ -2.76$ &   39   & $ -6.82$   \\  
 5516.78  &  MnI  &   2.18 & $ -1.85$ &  122   & $ -6.96$   \\  
 5537.76  &  MnI  &   2.19 & $ -2.21$ &   36   & $ -7.39$   \\  
 6013.48  &  MnI  &   3.07 & $ -0.25$ &  167   & $ -7.14$   \\  
 6016.64  &  MnI  &   3.07 & $ -0.24$ &  153   & $ -7.27$   \\  
 6021.79  &  MnI  &   3.08 & $  0.03$ &  144   & $ -7.61$   \\  
 6440.93  &  MnI  &   3.77 & $ -1.52$ &   06   & $ -7.10$   \\  
 5002.79  &  FeI  &   3.40 & $ -1.58$ &  206  &  $ -4.63$   \\  
 5028.13  &  FeI  &   3.57 & $ -1.10$ &  218  &  $ -4.75$   \\  
 5044.21  &  FeI  &   2.85 & $ -2.15$ &  161  &  $ -5.19$   \\  
 5054.64  &  FeI  &   3.64 & $ -2.14$ &   89  &  $ -4.89$   \\  
 5133.69  &  FeI  &   4.18 & $  0.14$ &  230  &  $ -5.17$   \\  
 5141.74  &  FeI  &   2.42 & $ -2.15$ &  242  &  $ -4.90$   \\  
 5162.27  &  FeI  &   4.18 & $  0.02$ &  224  &  $ -5.13$   \\  
 5228.41  &  FeI  &   4.22 & $ -1.29$ &  153  &  $ -4.51$   \\  
 5236.19  &  FeI  &   4.19 & $ -1.72$ &   97  &  $ -4.62$   \\  
 5243.78  &  FeI  &   4.26 & $ -1.15$ &  123  &  $ -4.88$   \\  
 5281.79  &  FeI  &   3.04 & $ -1.02$ &  247  &  $ -5.26$   \\  
 5288.53  &  FeI  &   3.69 & $ -1.67$ &  139  &  $ -4.89$   \\  
 5293.97  &  FeI  &   4.14 & $ -1.87$ &   96  &  $ -4.53$   \\  
 5302.30  &  FeI  &   3.28 & $ -0.88$ &  250  &  $ -5.06$   \\  
 5315.07  &  FeI  &   4.37 & $ -1.55$ &   97  &  $ -4.58$   \\  
 5321.11  &  FeI  &   4.43 & $ -1.44$ &   98  &  $ -4.61$   \\  
 5322.04  &  FeI  &   2.28 & $ -3.03$ &  145  &  $ -5.19$   \\  
 5329.99  &  FeI  &   4.08 & $ -1.30$ &  139  &  $ -4.81$   \\  
 5358.12  &  FeI  &   3.30 & $ -3.37$ &   25  &  $ -4.82$   \\  
 5364.87  &  FeI  &   4.45 & $  0.22$ &  223  &  $ -5.04$   \\  
 5365.40  &  FeI  &   3.57 & $ -1.28$ &  175  &  $ -5.09$   \\  
 5373.71  &  FeI  &   4.47 & $ -0.86$ &   91  &  $ -5.20$   \\  
 5379.57  &  FeI  &   3.69 & $ -1.48$ &  119  &  $ -5.26$   \\  
 5383.37  &  FeI  &   4.31 & $  0.50$ &  241  &  $ -5.30$   \\  
 5386.34  &  FeI  &   4.15 & $ -1.77$ &   40  &  $ -5.19$   \\  
 5391.46  &  FeI  &   4.15 & $ -0.94$ &  175  &  $ -4.74$   \\  
 5398.29  &  FeI  &   4.45 & $ -0.67$ &  112  &  $ -5.25$   \\  
 5400.50  &  FeI  &   4.37 & $ -0.16$ &  202  &  $ -4.98$   \\  
 5410.91  &  FeI  &   4.47 & $  0.28$ &  225  &  $ -5.06$   \\  
 5417.03  &  FeI  &   4.42 & $ -1.68$ &   34  &  $ -5.07$   \\  
 5441.32  &  FeI  &   4.31 & $ -1.73$ &   63  &  $ -4.79$   \\  
 5445.04  &  FeI  &   4.39 & $ -0.02$ &  203  &  $ -5.11$   \\  
 5452.12  &  FeI  &   3.64 & $ -2.86$ &   51  &  $ -4.57$   \\  
 5464.29  &  FeI  &   4.14 & $ -1.72$ &   86  &  $ -4.78$   \\  
 5522.46  &  FeI  &   4.21 & $ -1.55$ &   49  &  $ -5.24$   \\  
 5536.59  &  FeI  &   2.83 & $ -3.81$ &   12  &  $ -5.29$   \\  
 5543.15  &  FeI  &   3.69 & $ -1.57$ &  168  &  $ -4.74$   \\  
 5543.94  &  FeI  &   4.22 & $ -1.14$ &  126  &  $ -4.93$   \\  
 5554.89  &  FeI  &   4.55 & $ -0.44$ &  188  &  $ -4.66$   \\  
 5557.95  &  FeI  &   4.47 & $ -1.28$ &  122  &  $ -4.53$   \\  
 5560.23  &  FeI  &   4.43 & $ -1.19$ &  106  &  $ -4.80$   \\  
 5563.60  &  FeI  &   4.19 & $ -0.99$ &  158  &  $ -4.83$   \\  
 5565.71  &  FeI  &   4.61 & $ -0.29$ &  178  &  $ -4.84$   \\  
 5567.40  &  FeI  &   2.61 & $ -2.80$ &  141  &  $ -5.09$   \\  
 5618.65  &  FeI  &   4.21 & $ -1.38$ &  129  &  $ -4.68$   \\  
 5633.97  &  FeI  &   4.99 & $ -0.27$ &  131  &  $ -4.85$   \\  
 5635.85  &  FeI  &   4.26 & $ -1.89$ &   66  &  $ -4.67$   \\  
 5638.27  &  FeI  &   4.22 & $ -0.87$ &  124  &  $ -5.23$   \\  
 5652.32  &  FeI  &   4.26 & $ -1.95$ &   43  &  $ -4.86$   \\  
 5653.89  &  FeI  &   4.39 & $ -1.64$ &   55  &  $ -4.89$   \\  
 5679.02  &  FeI  &   4.65 & $ -0.92$ &  107  &  $ -4.81$   \\  
 5686.53  &  FeI  &   4.55 & $ -0.63$ &  166  &  $ -4.70$   \\  
 5717.85  &  FeI  &   4.28 & $ -1.13$ &  116  &  $ -4.96$   \\  
 5720.89  &  FeI  &   4.55 & $ -1.95$ &   40  &  $ -4.57$   \\  
 5738.22  &  FeI  &   4.22 & $ -2.34$ &   42  &  $ -4.54$   \\  
 5741.86  &  FeI  &   4.26 & $ -1.73$ &   51  &  $ -5.00$   \\  
 5752.04  &  FeI  &   4.55 & $ -0.99$ &   77  &  $ -5.14$   \\  
 5753.12  &  FeI  &   4.26 & $ -0.76$ &  183  &  $ -4.75$   \\  
 5775.09  &  FeI  &   4.22 & $ -1.23$ &   94  &  $ -5.13$   \\  
 5778.47  &  FeI  &   2.59 & $ -3.59$ &   76  &  $ -4.86$   \\  
 5793.93  &  FeI  &   4.22 & $ -1.70$ &   41  &  $ -5.19$   \\  
 5806.73  &  FeI  &   4.61 & $ -1.05$ &  124  &  $ -4.60$   \\  
 5807.79  &  FeI  &   3.29 & $ -3.41$ &   39  &  $ -4.60$   \\  
 5809.25  &  FeI  &   3.88 & $ -1.84$ &  103  &  $ -4.84$   \\  
 5826.64  &  FeI  &   4.28 & $ -2.94$ &   06  &  $ -4.80$   \\  
 5827.89  &  FeI  &   3.28 & $ -3.41$ &   41  &  $ -4.58$   \\  
 5833.93  &  FeI  &   2.61 & $ -3.66$ &   54  &  $ -4.98$   \\  
 5852.19  &  FeI  &   4.55 & $ -1.33$ &  103  &  $ -4.57$   \\  
 5855.13  &  FeI  &   4.61 & $ -1.76$ &   52  &  $ -4.55$   \\  
 5856.08  &  FeI  &   4.29 & $ -1.64$ &   71  &  $ -4.84$   \\  
 5859.61  &  FeI  &   4.55 & $ -0.60$ &  167  &  $ -4.73$   \\  
 5862.36  &  FeI  &   4.55 & $ -0.38$ &  132  &  $ -5.27$   \\  
 5873.21  &  FeI  &   4.26 & $ -2.14$ &   53  &  $ -4.57$   \\  
 5883.84  &  FeI  &   3.96 & $ -1.36$ &  170  &  $ -4.66$   \\  
 5902.52  &  FeI  &   4.59 & $ -1.81$ &   23  &  $ -4.95$   \\  
 5916.25  &  FeI  &   2.45 & $ -2.99$ &  180  &  $ -4.80$   \\  
 5927.80  &  FeI  &   4.65 & $ -1.09$ &   95  &  $ -4.76$   \\  
 5930.17  &  FeI  &   4.65 & $ -0.23$ &  205  &  $ -4.62$   \\  
 5934.66  &  FeI  &   3.93 & $ -1.17$ &  128  &  $ -5.26$   \\  
 5952.75  &  FeI  &   3.98 & $ -1.44$ &   93  &  $ -5.21$   \\  
 5984.80  &  FeI  &   4.73 & $ -0.31$ &  180  &  $ -4.69$   \\  
 5987.06  &  FeI  &   4.79 & $ -0.59$ &  137  &  $ -4.72$   \\  
 6003.03  &  FeI  &   3.88 & $ -1.12$ &  140  &  $ -5.27$   \\  
 6020.17  &  FeI  &   4.61 & $ -0.27$ &  193  &  $ -4.75$   \\  
 6024.07  &  FeI  &   4.55 & $ -0.12$ &  159  &  $ -5.30$   \\  
 6055.99  &  FeI  &   4.73 & $ -0.46$ &  112  &  $ -5.17$   \\  
 6082.72  &  FeI  &   2.22 & $ -3.57$ &  140  &  $ -4.85$   \\  
 6094.42  &  FeI  &   4.65 & $ -1.94$ &   35  &  $ -4.55$   \\  
 6096.69  &  FeI  &   3.98 & $ -1.93$ &   83  &  $ -4.82$   \\  
 6105.15  &  FeI  &   4.55 & $ -2.05$ &   18  &  $ -4.89$   \\  
 6127.91  &  FeI  &   4.14 & $ -1.69$ &   93  &  $ -4.79$   \\  
 6157.73  &  FeI  &   4.08 & $ -1.26$ &  132  &  $ -4.97$   \\  
 6180.22  &  FeI  &   2.73 & $ -2.78$ &  129  &  $ -5.12$   \\  
 6188.04  &  FeI  &   3.94 & $ -1.72$ &   87  &  $ -5.05$   \\  
 6200.32  &  FeI  &   2.61 & $ -2.44$ &  181  &  $ -5.17$   \\  
 6213.43  &  FeI  &   2.22 & $ -2.66$ &  210  &  $ -5.20$   \\  
 6215.15  &  FeI  &   4.19 & $ -1.44$ &  150  &  $ -4.51$   \\  
 6226.77  &  FeI  &   3.88 & $ -2.22$ &   63  &  $ -4.84$   \\  
 6229.23  &  FeI  &   2.85 & $ -2.97$ &  108  &  $ -4.95$   \\  
 6232.64  &  FeI  &   3.65 & $ -1.33$ &  175  &  $ -5.04$   \\  
 6246.32  &  FeI  &   3.60 & $ -0.96$ &  244  &  $ -4.84$   \\  
 6254.26  &  FeI  &   2.28 & $ -2.48$ &  249  &  $ -4.96$   \\  
 6271.29  &  FeI  &   3.33 & $ -2.95$ &   90  &  $ -4.53$   \\  
 6301.50  &  FeI  &   3.65 & $ -0.59$ &  241  &  $ -5.20$   \\  
 6311.51  &  FeI  &   2.83 & $ -3.23$ &   67  &  $ -5.05$   \\  
 6336.84  &  FeI  &   3.69 & $ -1.05$ &  239  &  $ -4.72$   \\  
 6353.84  &  FeI  &   0.91 & $ -6.48$ &   17  &  $ -4.85$   \\  
 6355.04  &  FeI  &   2.85 & $ -2.42$ &  177  &  $ -4.96$   \\  
 6380.75  &  FeI  &   4.19 & $ -1.40$ &  112  &  $ -4.88$   \\  
 6392.55  &  FeI  &   2.28 & $ -4.03$ &   71  &  $ -4.89$   \\  
 6408.02  &  FeI  &   3.69 & $ -1.00$ &  221  &  $ -4.94$   \\  
 6411.65  &  FeI  &   3.65 & $ -0.82$ &  211  &  $ -5.25$   \\  
 6419.98  &  FeI  &   4.73 & $ -0.24$ &  145  &  $ -5.11$   \\  
 6518.38  &  FeI  &   2.83 & $ -2.75$ &  173  &  $  -4.6$9  \\  
 6581.22  &  FeI  &   1.49 & $ -4.86$ &  139  &  $  -4.5$4  \\  
 6593.88  &  FeI  &   2.43 & $ -2.42$ &  206  &  $  -5.2$5  \\  
 6597.61  &  FeI  &   4.80 & $ -1.07$ &   90  &  $  -4.6$9  \\  
 6608.03  &  FeI  &   2.28 & $ -4.03$ &   41  &  $  -5.2$1  \\  
 6609.12  &  FeI  &   2.56 & $ -2.69$ &  185  &  $  -5.0$1  \\  
 6627.56  &  FeI  &   4.55 & $ -1.68$ &   60  &  $  -4.6$5  \\  
 6646.98  &  FeI  &   2.61 & $ -3.99$ &   50  &  $  -4.7$5  \\  
 6653.88  &  FeI  &   4.15 & $ -2.52$ &   26  &  $  -4.7$2  \\  
 6703.57  &  FeI  &   2.76 & $ -3.16$ &  112  &  $  -4.8$6  \\  
 6705.10  &  FeI  &   4.61 & $ -1.28$ &   98  &  $  -4.6$4  \\  
 6710.31  &  FeI  &   1.49 & $ -4.88$ &   95  &  $  -4.8$4  \\  
 6715.41  &  FeI  &   4.61 & $ -1.64$ &   51  &  $  -4.7$2  \\  
 6726.67  &  FeI  &   4.61 & $ -1.12$ &  114  &  $  -4.6$7  \\  
 6733.16  &  FeI  &   4.64 & $ -1.58$ &   42  &  $  -4.8$6  \\  
 6737.98  &  FeI  &   4.56 & $ -1.75$ &   59  &  $  -4.5$8  \\  
 6739.54  &  FeI  &   1.56 & $ -4.95$ &   54  &  $  -5.0$4  \\  
 6750.15  &  FeI  &   2.42 & $ -2.62$ &  189  &  $  -5.2$2  \\  
 6752.72  &  FeI  &   4.64 & $ -1.36$ &   63  &  $  -4.8$4  \\  
 6806.85  &  FeI  &   2.73 & $ -3.21$ &  110  &  $  -4.8$7  \\  
 6810.28  &  FeI  &   4.61 & $ -1.12$ &   86  &  $  -4.9$1  \\  
 6837.00  &  FeI  &   4.59 & $ -1.81$ &   28  &  $  -4.9$0  \\  
 6843.67  &  FeI  &   4.55 & $ -0.93$ &  108  &  $  -4.9$8  \\  
 6855.16  &  FeI  &   4.56 & $ -0.63$ &  172  &  $  -4.7$2  \\  
 6858.16  &  FeI  &   4.61 & $ -1.06$ &   80  &  $  -5.0$2  \\  
 5132.66  &  FeII &   2.81 & $ -4.18$ &  103  &  $ -4.53$   \\  
 5197.56  &  FeII &   3.23 & $ -2.10$ &  213  &  $ -4.95$   \\  
 5534.83  &  FeII &   3.24 & $ -2.93$ &  111  &  $ -5.20$   \\  
 5991.37  &  FeII &   3.15 & $ -3.74$ &  114  &  $ -4.48$   \\  
 6084.10  &  FeII &   3.20 & $ -3.98$ &   55  &  $ -4.81$   \\  
 6113.33  &  FeII &   3.22 & $ -4.31$ &   38  &  $ -4.69$   \\  
 6149.25  &  FeII &   3.89 & $ -2.92$ &   89  &  $ -4.71$   \\  
 6238.38  &  FeII &   3.89 & $ -2.87$ &  118  &  $ -4.47$   \\  
 6247.55  &  FeII &   3.89 & $ -2.51$ &  126  &  $ -4.73$   \\  
 6369.46  &  FeII &   2.89 & $ -4.36$ &   39  &  $ -4.99$   \\  
 6383.72  &  FeII &   5.55 & $ -2.27$ &   10  &  $ -4.95$   \\  
 6416.92  &  FeII &   3.89 & $ -2.85$ &   56  &  $ -5.16$   \\  
 6432.68  &  FeII &   2.89 & $ -3.74$ &   90  &  $ -5.03$   \\  
 7479.69  &  FeII &   3.89 & $ -3.88$ &   11  &  $ -5.05$   \\  
 7711.71  &  FeII &   3.90 & $ -2.74$ &   91  &  $ -4.89$   \\  
 5212.70  &  CoI  &   3.51 & $ -0.11$ &   76  &  $ -7.33$   \\  
 5342.71  &  CoI  &   4.02 & $  0.36$ &   81  &  $ -7.17$   \\  
 5352.05  &  CoI  &   3.58 & $  0.06$ &   73  &  $ -7.46$   \\  
 5359.20  &  CoI  &   4.15 & $ -0.09$ &   19  &  $ -7.38$   \\  
 5523.30  &  CoI  &   2.33 & $ -1.85$ &   34  &  $ -7.48$   \\  
 5530.78  &  CoI  &   1.71 & $ -2.06$ &   80  &  $ -7.53$   \\  
 5647.23  &  CoI  &   2.28 & $ -1.56$ &   34  &  $ -7.84$   \\  
 6249.50  &  CoI  &   2.04 & $ -2.41$ &   69  &  $ -6.92$   \\  
 6632.44  &  CoI  &   2.28 & $ -2.00$ &   62  &  $ -7.14$   \\  
 6814.95  &  CoI  &   1.95 & $ -1.90$ &  125  &  $ -7.14$   \\  
 5010.96  &  NiI  &   3.63 & $ -0.87$ &  105  &  $ -6.29$   \\  
 5017.58  &  NiI  &   3.54 & $ -0.08$ &  186  &  $ -6.41$   \\  
 5032.75  &  NiI  &   3.90 & $ -1.27$ &   34  &  $ -6.33$   \\  
 5035.37  &  NiI  &   3.63 & $  0.29$ &  208  &  $ -6.43$   \\  
 5084.08  &  NiI  &   3.68 & $  0.03$ &  184  &  $ -6.38$   \\  
 5094.42  &  NiI  &   3.83 & $ -1.08$ &   63  &  $ -6.25$   \\  
 5102.97  &  NiI  &   1.68 & $ -2.62$ &  159  &  $ -6.43$   \\  
 5155.77  &  NiI  &   3.90 & $ -0.09$ &  163  &  $ -6.24$   \\  
 5176.57  &  NiI  &   3.90 & $ -0.44$ &  102  &  $ -6.46$   \\  
 5578.72  &  NiI  &   1.68 & $ -2.64$ &  179  &  $ -6.31$   \\  
 5847.00  &  NiI  &   1.68 & $ -3.21$ &   81  &  $ -6.55$   \\  
 6086.29  &  NiI  &   4.27 & $ -0.53$ &   85  &  $ -6.13$   \\  
 6175.42  &  NiI  &   4.09 & $ -0.53$ &   90  &  $ -6.30$   \\  
 6327.60  &  NiI  &   1.68 & $ -3.15$ &   90  &  $ -6.59$   \\  
 6378.26  &  NiI  &   4.15 & $ -0.89$ &   64  &  $ -6.12$   \\  
 6586.33  &  NiI  &   1.95 & $ -2.81$ &  135  &  $ -6.26$   \\  
 6643.64  &  NiI  &   1.68 & $ -2.30$ &  175  &  $ -6.82$   \\  
 6767.77  &  NiI  &   1.83 & $ -2.17$ &  168  &  $ -6.82$   \\  
 6772.36  &  NiI  &   3.66 & $ -0.98$ &   83  &  $ -6.45$   \\  
 7110.91  &  NiI  &   1.93 & $ -2.98$ &  111  &  $ -6.33$   \\  
 7122.24  &  NiI  &   3.54 & $  0.04$ &  253  &  $ -6.19$   \\  
 5105.55  &  CuI  &   1.39 & $ -1.51$ &  203  &  $ -8.57$   \\  
 5218.21  &  CuI  &   3.82 & $  0.27$ &  120  &  $ -8.20$   \\  
 4722.16  &  ZnI  &   4.03 & $ -0.39$ &  138  &  $ -8.02$   \\  
 4810.54  &  ZnI  &   4.08 & $ -0.17$ &  158  &  $ -7.98$   \\  
 5630.14  &  YI   &   1.36 & $  0.15$ &   07  &  $-10.01$   \\  
 6023.41  &  YI   &   0.00 & $ -1.85$ &   04  &  $-10.05$   \\  
 6222.61  &  YI   &   0.00 & $ -1.69$ &   02  &  $-10.43$   \\  
 6687.50  &  YI   &   0.00 & $ -0.43$ &   17  &  $-10.76$   \\  
 5087.42  &  YII  &   1.08 & $ -0.17$ &  139  &  $-10.70$   \\  
 5200.41  &  YII  &   0.99 & $ -0.57$ &  139  &  $-10.42$   \\  
 5289.82  &  YII  &   1.03 & $ -1.85$ &   28  &  $-10.24$   \\  
 5728.89  &  YII  &   1.84 & $ -1.12$ &   24  &  $-10.14$   \\  
 5853.67  &  BaII &   0.60 & $ -1.00$ &  186  &  $-10.68$   \\  
 6141.71  &  BaII &   0.70 & $ -0.08$ &  270  &  $-10.74$   \\  
 6496.90  &  BaII &   0.60 & $ -0.38$ &  299  &  $-10.39$   \\  
 6262.25  &  LaII &   0.40 & $ -1.45$ &   43  &  $-10.93$   \\  
 6320.41  &  LaII &   0.17 & $ -1.42$ &   32  &  $-11.39$   \\  
 6390.48  &  LaII &   0.32 & $ -1.49$ &   23  &  $-11.31$   \\  
 5330.58  &  CeII &   0.86 & $ -0.23$ &   27  &  $-11.05$   \\  
 5610.24  &  CeII &   1.04 & $  0.00$ &   49  &  $-10.76$   \\  
 6043.38  &  CeII &   1.20 & $ -0.17$ &   08  &  $-11.32$   \\  
 5219.03  &  PrII &   0.79 & $ -0.24$ &   09  &  $-11.95$   \\  
 5322.82  &  PrII &   0.48 & $ -0.54$ &   24  &  $-11.55$   \\  
 5234.21  &  NdII &   0.55 & $ -0.33$ &   92  &  $-10.84$   \\  
 5293.17  &  NdII &   0.82 & $ -0.06$ &  122  &  $-10.53$   \\  
 5311.48  &  NdII &   0.99 & $ -0.42$ &   26  &  $-11.00$   \\  
 5485.71  &  NdII &   1.26 & $ -0.12$ &   38  &  $-10.79$   \\  
 5740.88  &  NdII &   1.16 & $ -0.55$ &   11  &  $-11.10$   \\  
 5842.39  &  NdII &   1.28 & $ -0.60$ &   08  &  $-11.05$   \\  
 6031.31  &  NdII &   1.28 & $ -0.70$ &   17  &  $-10.62$   \\  
 6437.64  &  EuII &   1.32 & $ -0.28$ &   15  &  $-11.82$   \\  
 6645.13  &  EuII &   1.38 & $  0.20$ &   24  &  $-12.01$   \\  
\end{supertabular}
\end{center}

\begin{table}					    
\caption{Chemical composition $\log \epsilon(X)$	    
      ($ \log \epsilon(H)=12.0$); ``n'' means the number of
      lines used, $\sigma$ -- the dispersion of the derived abundance for the given
      number of lines; model parameters are indicated under the spectrum number;
      abundances in the solar photosphere are taken from [\cite{Grevesse}].}
\bigskip						    
\begin{tabular}{lc@{\qquad}|lcrccc}			    
\hline							    
& Sun  &\multicolumn{5}{c}{\qquad IRAS\,20508\,+\,2011\qquad} & \\
  & & &  \multicolumn{4}{c}{\qquad 4800\,K, 1.5, 4.0\,km/s \qquad\qquad} 
                      &   \\
\cline{4-7}
 E & $\log \epsilon(E)$ &  $X$ \qquad & $\log \epsilon(X)$
      & n &\qquad $\sigma$ \qquad
  & $[X/Fe]_{\odot}$ \\
\hline
C      &8.55 & CI  &{\it 9.07}&4&0.24 &$ {\it +0.88}$ &  \\
N      &7.97 & NI  &{\it 9.33}&1&     &$ {\it +1.72}$ &  \\
O      &8.87 & OI  & 10.30&  3 & 0.22 &$ +1.79$ &   \\
Na     &6.33 & NaI &  6.55&  4 & 0.30 &$ +0.58$ & \\
Mg     &7.58 & MgI &  6.87&  3 & 0.20 &$ -0.35$ &  \\
Al     &6.47 & AlI &  6.43&  3 & 0.11 &$ +0.32$ &  \\
Si     &7.55 & SiI &  7.27& 14 & 0.20 &$ +0.08$ & \\
       &     & SiII&{\it 7.52} &  1 & &$ {\it +0.33}$ & \\
S      &7.21 & Si  &  7.59&  5 & 0.37 &$ +0.74$ & \\
Ca     &6.36 & CaI &  6.04& 18 & 0.22 &$ +0.04$ & \\
Sc     &3.17 & ScII&  2.59&  7 & 0.21 &$ -0.20$ & \\
Ti     &5.02 & TiI &  4.51& 35 & 0.21 &$ -0.15$ &  \\
       &     & TiII&  4.38& 8  & 0.25 &$ -0.28$ & \\
V      &4.00 & VI  &  3.50& 17 & 0.28 &$ -0.14$ & \\
       &     & VII &  3.60&  3 & 0.21 &$ -0.04$ &  \\
Cr     &5.67 & CrI &  5.54& 16 & 0.18 &$ +0.23$ & \\
       &     & CrII&  5.41&  6 & 0.32 &$ +0.10$ & \\
Mn     &5.39 & MnI &  4.81&  12& 0.30 &$ -0.22$ & \\
Fe     &7.50 & FeI &  7.11&138 & 0.23 &$ +0.03$ & \\
       &     & FeII&  7.16& 15 & 0.24 &$ -0.02$ & \\
Co     &4.92 & CoI &  4.66& 10 & 0.26 &$ +0.10$ &  \\  
Ni     &6.25 & NiI &  5.61& 21 & 0.19 &$ -0.28$ &  \\
Cu     &4.21 & CuI &  3.62&  2 &      &$ -0.23$ &  \\
Zn     &4.60 & ZnI &  4.00&  2 &      &$ -0.24$ &  \\
Y      &2.24 & YI  &  1.70&  4 & 0.35 &$ -0.18$ &  \\
       &     & YII &  1.62&  4 & 0.25 &$ -0.26$ &  \\
Zr     &2.60 & ZrI &  1.57&  2 &      &$ -0.67$ &  \\
Ba     &2.13 & BaII&  1.32&  1 &      &$ -0.45$ &   \\
       &     &     &  1.40&  3 & 0.19 &$ -0.37$ &   \\
La     &1.22 & LaII&  0.79&  3 & 0.25 &$ -0.07$ & \\
Ce     &1.55 & CeII&  0.96&  3 & 0.28 &$ -0.28$ &   \\
Pr     &0.71 & PrII&  0.25&  2 &      &$ -0.10$ & \\
Nd     &1.50 & NdII&  1.15&  7 & 0.22 &$ +0.01$ & \\
Eu     &0.51 & EuII&  0.08&  2 &      &$ -0.07$ &  \\
\hline
\end{tabular}
\end{table}

\end{document}